\documentclass[11pt,a4paper]{article}

\usepackage[T1]{fontenc}
\usepackage[utf8]{inputenc}
\usepackage[margin=2.5cm]{geometry}
\usepackage{amsmath,amssymb}
\usepackage{graphicx}
\usepackage{booktabs}
\usepackage{array}
\usepackage{float}
\usepackage[font=small,labelfont=bf]{caption}
\usepackage{subcaption}
\usepackage{enumitem}
\usepackage{cite}
\usepackage[hidelinks]{hyperref}
\usepackage[capitalise,noabbrev]{cleveref}

\newcommand{\That}{\hat{T}_\theta}
\newcommand{\Lpde}{\mathcal{L}_{\mathrm{PDE}}}
\newcommand{\stage}[1]{\texttt{#1}}
\newcommand{\fwd}{\stage{forward}}
\newcommand{\firstgrad}{\stage{first\_grad}}
\newcommand{\secondgrad}{\stage{second\_grad}}
\newcommand{\paramgrad}{\stage{param\_grad}}
\newcommand{\sci}[2]{$#1\times10^{#2}$}
\newcommand{\dqubit}{\texttt{default.qubit}}
\newcommand{\lqubit}{\texttt{lightning.qubit}}

\graphicspath{{./}}

\usepackage{authblk}

\title{Fewer Parameters, Longer Runtimes: The Wall-Clock Cost of\\
Hybrid Quantum Physics-Informed Neural Networks}

\author[a]{Leyan Lyu\thanks{\texttt{llyu@terpmail.umd.edu}}}
\author[a,b]{Shabnam Jabeen\thanks{\texttt{jabeen@umd.edu}}}

\affil[a]{Department of Physics, University of Maryland, College Park, MD 20742, U.S.A.}
\affil[b]{National Quantum Lab, University of Maryland, College Park, MD 20742, U.S.A.}

\date{}

\begin{document}
\maketitle

\begin{abstract}

While hybrid quantum-classical physics-informed neural networks (QCPINNs) are often described as efficient because they have fewer trainable parameters than classical PINNs, parameter count does not measure wall-clock cost. To examine this gap on a one-dimensional seawater temperature diffusion benchmark, with the quantum circuits simulated on a CPU, we time the four cumulative stages a PINN needs (the forward pass, the first- and second-order input gradients, and the PDE-loss parameter gradient) for a classical PINN, four hybrid configurations that differ in qubit count and circuit depth, and size-matched classical controls that share the hybrids' classical layers. We find that the hybrids are 14--56$\times$ slower than the classical PINN for the forward pass and 10--97$\times$ slower for the PDE-loss parameter gradient, and even against the size-matched controls, they remain 9--85$\times$ slower. Further, over the tested ranges, runtime increased approximately exponentially with qubit count, approximately linearly with circuit depth, and sub-linearly with batch size, while for the circuit-only forward pass, a second simulator backend was 10--165$\times$ slower under the same calling pattern. Although the hybrids have 38--44\% fewer parameters than the classical PINN, this reduction comes from their smaller classical layers, as the quantum circuit accounts for at most 2.5\% of each hybrid's parameters. Yet, parameter count failed to predict runtime even among the classical networks, since the size-matched controls, despite having fewer parameters than the classical PINN, run slower than it. Our conclusion is that fewer parameters did not mean lower cost: every hybrid took at least ten times longer than the classical PINN at every stage, and the gap grew quickly with the number of qubits.

\end{abstract}

\section{Introduction}\label{sec:intro}

Many problems in physics and engineering are described by partial differential equations (PDEs), equations that relate how a quantity, such as temperature, changes in space and in time. Physics-informed neural networks (PINNs)~\cite{raissi2019} solve forward and inverse PDE problems with a neural network. A neural network, at its core, is a flexible mathematical function containing many adjustable numbers, or trainable parameters, that are tuned during training. A PINN, more specifically, is trained so that its output obeys the governing equation, the boundary conditions, and, for inverse problems (where an unknown physical constant is inferred from measurements), a set of observations, all of which are written into the single quantity that training minimizes, known as the loss.

We study the one-dimensional temperature diffusion benchmark of Han et al.~\cite{han2026}, derived from the vertical diffusion of seawater temperature, under Dirichlet, Neumann, and Robin boundary conditions (\cref{sec:problem}). Fast, differentiable surrogates for such equations could also serve energy-aware control of engineered water bodies, such as swimming pools or other managed basins, where a model that returns temperature and its gradients cheaply could inform heating, circulation, or shading decisions within an optimal-control loop.

Hybrid quantum-classical PINNs (QCPINNs) replace part of the network with a variational quantum circuit (VQC), a small quantum computation whose settings are tuned during training like ordinary network weights (\cref{sec:vqc}). As reported by Farea et al.~\cite{farea2025}, QCPINNs can reach accuracy comparable to classical PINNs while using far fewer trainable parameters, and for this reason, parameter count has become a prominent efficiency metric in the studies discussed here.

\paragraph{Gap.} Yet, while a smaller parameter count might seem to promise a cheaper model, it does not by itself guarantee one. Wall-clock time, the time a computer takes to finish a calculation, is rarely the headline metric, and where it is reported, it appears as a single aggregate figure (\cref{sec:related}). In the simulator-based setting considered here, VQCs run on ordinary computers that simulate the quantum circuit, and the simulation cost grows with the circuit's width and depth (\cref{sec:vqc}). Further, training a PINN requires first- and second-order derivatives of the network's output with respect to its inputs, which cost more than a plain evaluation of the network and must pass through the simulated circuit. The question remains, however: once the circuit must be simulated and differentiated, how much of the reported saving in parameters carries over to actual runtime?

\paragraph{Research questions.}
\begin{description}[leftmargin=3em,style=nextline]
  \item[RQ1] How much wall-clock overhead do hybrid models incur in each of the four cumulative stages, relative to the classical baseline and to size-matched classical controls, and does this overhead follow parameter count?
  \item[RQ2] How does this cost scale with qubit count, circuit depth, batch size, and simulator backend?
\end{description}
A single-seed training check (\cref{sec:training}) supplements these questions: it tests whether the hybrids train on the benchmark and shows how their per-evaluation cost shaped the training schedule used here.

\paragraph{Contributions.}
\begin{enumerate}
  \item Stage-wise timing of four cumulative computations (the forward pass, the first-order input gradient, the second-order input gradient, and the PDE-loss parameter gradient) for a classical PINN, four hybrid configurations, and size-matched classical controls that share the hybrids' classical layers.
  \item A sweep of qubit count, circuit depth, batch size, and simulator backend.
  \item A parameter-count breakdown showing that the tested models'
parameter counts do not track their wall-clock costs.
  \item A single-seed training sanity check across six scenarios of one PDE family, showing how per-evaluation cost influenced our choice of optimizer.
\end{enumerate}

\paragraph{Outline.} \Cref{sec:related} reviews related work. \Cref{sec:background} defines the test problem, the PINN loss, the four timed stages, and the quantum-circuit basics used throughout. \Cref{sec:models} describes the classical baseline, the four hybrid configurations and the size-matched classical controls. \Cref{sec:protocol} details the timing protocol. \Cref{sec:timing,sec:scaling} report stage-wise wall-clock cost and its scaling, \cref{sec:params} compares it with parameter count, and \cref{sec:training} reports the training check. \Cref{sec:discussion,sec:conclusion} discuss the results and conclude.

\section{Related work}\label{sec:related}

Hybrid quantum PINNs have been proposed for fluid flow, wave, Helmholtz, and Klein--Gordon problems~\cite{farea2025,leong2025}, reservoir seepage~\cite{rao2025}, hydrological forecasting~\cite{hewage2026}, and Maxwell's equations~\cite{chen2026}. Theoretical work addresses universal approximation and the mitigation of barren plateaus~\cite{lantigua2025}; a barren plateau is a training landscape so flat that the gradients guiding training almost vanish.

Throughout this literature, the main efficiency claim remains parameter count. QCPINNs are reported to reach comparable accuracy with 10--30\% of the parameters of a classical PINN~\cite{farea2025}, quantum and hybrid models to outperform a classical benchmark with fewer parameters in some cases~\cite{leong2025,trahan2024}, and one study reports about 44\% fewer parameters together with convergence in roughly three times fewer training epochs, an epoch count rather than a runtime comparison~\cite{hewage2026}. An epoch is one full pass of training over the data; it does not specify how much computation that pass requires.

While these studies do acknowledge cost, they usually report it as one aggregate figure. As noted by Farea et al.~\cite{farea2025}, time per training iteration increases 15- to 40-fold relative to classical PINNs, at similar memory. Leong et al.~\cite{leong2025} note that simulating the quantum layers imposes an overhead of at least two orders of magnitude for the same number of trainable parameters, but compare training efficiency through epochs rather than wall-clock time. Trahan et al.~\cite{trahan2024} report that a noiseless quantum PINN on a 2D Poisson problem took nearly four times as long as a classical PINN with a similar number of parameters.

Since these figures differ widely and depend on hardware, simulator, circuit, and batch size, comparing them directly is difficult. Yet, among the studies reviewed here, none breaks the cost down into the four cumulative stages a PINN needs (forward evaluation, first- and second-order input gradients, and the parameter gradient), or varies qubit count, depth, batch size, and backend in a controlled way. On the simulation side, Chen et al.~\cite{chen2026} develop a GPU-accelerated simulation library for computing circuit outputs and their derivatives to make QPINN training practical. This paper addresses that gap with a controlled, stage-wise timing study alongside the parameter count.

\section{Background and key concepts}\label{sec:background}

\subsection{Test problem and PINN loss}\label{sec:problem}

We consider the diffusion equation
\begin{equation}\label{eq:pde}
  \frac{\partial T}{\partial t} = k_z \frac{\partial^2 T}{\partial z^2},
  \qquad z \in [0, 2],\; t \in [0, 5],
\end{equation}
with diffusivity $k_z = 0.05$ in forward problems and $k_z$ estimated from data in inverse problems. Here $T$ is temperature, $z$ is depth, and $t$ is time. In other words, \cref{eq:pde} states that the rate of change of temperature in time (left side) equals the diffusivity times its curvature in depth (right side). As such, checking whether a network's prediction obeys this equation means computing the derivatives each side needs, which \cref{sec:stages} describes and \cref{sec:timing} times.

Let $\That(z,t)$ be the network output with trainable parameters $\theta$ and input $x = (z,t)$. The PINN loss combines four mean-squared terms:
\begin{equation}\label{eq:loss}
  \mathcal{L}(\theta) = w_{\mathrm{PDE}}\,\Lpde + \mathcal{L}_{\mathrm{BC}} + \mathcal{L}_{\mathrm{IC}} + \mathcal{L}_{\mathrm{data}}.
\end{equation}
These are the PDE residual at interior collocation points (how far the prediction is from satisfying \cref{eq:pde}), the boundary-condition residual, the misfit to the initial condition and, for inverse problems only, the misfit to observations. Collocation points refer to the sample locations where the equation is evaluated. The boundary, initial, and data terms are assigned unit weight. The PDE term has weight $w_{\mathrm{PDE}} = 1$ for the Dirichlet and Neumann problems and $w_{\mathrm{PDE}} = 10$ for the Robin problems, which needed both this weight and a larger step budget to converge (\cref{sec:training}). The weight does not change which derivatives are computed, so it does not affect the timings in \cref{sec:timing}.

\paragraph{Initial and boundary conditions.} A boundary condition fixes what happens at the edges of the domain: while a Dirichlet condition fixes the temperature itself, a Neumann condition fixes its slope, i.e.\ the heat flow across the boundary, and a Robin condition fixes a combination of the two. Closed-form solutions exist for all three types, and \cref{tab:bc} lists the conditions imposed. The initial condition is the exact solution at $t = 0$. For the Dirichlet problem,
the exact solution vanishes at both boundaries, so $\mathcal{L}_{\mathrm{BC}}$ penalizes $\That(0,t)$ and $\That(2,t)$ directly. The Neumann and Robin conditions involve $\partial\That/\partial z$ at the boundaries and therefore already require a first-order input gradient (\cref{sec:stages}).

\begin{table}[ht]
\centering
\caption{Initial and boundary conditions of the three benchmark problems. Each boundary condition holds at $z = 0$ and $z = 2$ for all $t \in [0,5]$.}
\label{tab:bc}
\begin{tabular}{lll}
\toprule
Type & Boundary condition & Initial condition $T(z,0)$ \\
\midrule
Dirichlet & $T = 0$ & $\sin\frac{\pi z}{2} + \sin \pi z$ \\
Neumann & $\partial T/\partial z = 0$ & $\cos\frac{\pi z}{2} + \cos \pi z$ \\
Robin & $\partial T/\partial z + hT = 0$, $h=1$ & $e^{-z} + \frac{\pi}{2}\cos\frac{\pi z}{2} - \sin\frac{\pi z}{2} + \pi\cos\pi z - \sin \pi z$ \\
\bottomrule
\end{tabular}
\end{table}

\paragraph{Training points.} We sample 2540 interior collocation points uniformly in $[0,2]\times[0,5]$, 40 points on each of the two boundaries, and 160 points at $t = 0$. Inverse problems additionally use 12 observations on a uniform $4\times3$ grid in $(z,t)$.

\paragraph{Relation to Han et al.} The governing equation, the domain, the diffusivity, the three boundary and initial conditions (with Robin coefficient $h = 1$), the numbers of collocation, boundary and initial points, the 12 inverse-problem observations and the classical network of \cref{sec:models} follow Han et al.~\cite{han2026}. The Adam phase, the step budgets and the Robin PDE weight $w_{\mathrm{PDE}} = 10$ were chosen for this study. Han et al.\ train with L-BFGS and unit loss weights by default; their sensitivity analysis found that raising $w_{\mathrm{PDE}}$ towards 10 generally reduced the error.

\subsection{Four cumulative timed stages}\label{sec:stages}

Training and evaluating a PINN involves four computations of increasing cost, each reusing the work of the one before it. The derivatives involved are computed by automatic differentiation~\cite{baydin2018,lu2021,raissi2019}: the software records every arithmetic step of the network's calculation and then applies the chain rule backward through that record. This backward sweep is called backpropagation. We use reverse-mode automatic differentiation in PyTorch~\cite{paszke2017}. Each reported stage time covers one complete call on a batch of points, with 2540 points in the main experiment.

\begin{enumerate}
  \item \fwd{} evaluates the network once, $\That(z,t)$, with no derivatives at all. (The stage name refers to this computation only; it is unrelated to the forward-problem versus inverse-problem distinction of \cref{sec:training}.)
  \item \firstgrad{} additionally computes the first-order input gradient $\nabla_x\That = (\partial\That/\partial z,\, \partial\That/\partial t)$: how the predicted temperature changes when depth or time changes slightly. It supplies the time derivative on the left of \cref{eq:pde} and the Neumann and Robin boundary residuals.
  \item \secondgrad{} adds the second-order input gradient $\partial^2\That/\partial z^2$, the diffusion term on the right of \cref{eq:pde}. It is obtained by differentiating $\partial\That/\partial z$ once more with respect to $z$, which requires a second backward pass through the record created by the first; differentiating through a gradient in this way is known as double backpropagation~\cite{drucker1992}. Only once this is available can the PDE residual be evaluated:
  \begin{equation}\label{eq:residual}
    r_\theta(z,t) = \frac{\partial\That}{\partial t} - k_z\frac{\partial^2\That}{\partial z^2}.
  \end{equation}
  The \secondgrad{} stage represents the computational cost of verifying whether a prediction satisfies the governing equation.

  \item \paramgrad{} adds the parameter gradient $\partial\Lpde/\partial\theta$, where $\Lpde = \frac{1}{N}\sum_i r_\theta(z_i,t_i)^2$ over the $N$ collocation points. This gradient informs the optimizer how to adjust each trainable parameter to minimize the loss. It is obtained by backpropagating through $r_\theta$ itself, through the record already built for the input gradients, one more time, this time to the network's weights $\theta$ rather than to its input $x$. It supplies the PDE-loss gradient used during training; it does not include the complete optimizer step.
\end{enumerate}

Because the stages are cumulative, \secondgrad{} necessarily includes the computations performed by \fwd{} and \firstgrad{}. The \secondgrad{}/\fwd{} ratio is therefore the full cost of the PDE residual relative to one forward pass, not the cost of the second derivative alone. \paramgrad{} accounts solely for the PDE term of $\mathcal{L}$; the boundary, initial-condition, and, in the case of inverse problems, data terms from \cref{eq:loss} introduce additional, less expensive backward passes that are not timed separately. Therefore, the total cost of a full training step is somewhat greater than that of \paramgrad{}.

\subsection{Variational quantum circuits}\label{sec:vqc}

A qubit is the quantum counterpart of a bit. A quantum circuit is a sequence of operations, called gates, applied to a set of qubits. A VQC acts on $n$ qubits and consists of an encoding that writes classical numbers into gate angles, followed by $L$ layers of single-qubit rotations with adjustable angles and entangling gates that link the qubits. Its outputs here are Pauli-$Z$ expectation values: averages of a measurement with outcomes $-1$ and $+1$, one real number per qubit between $-1$ and $+1$~\cite{benedetti2019,cerezo2021}. The trainable quantum parameters are the rotation angles; $n$ is the width and $L$ the depth of the circuit. The encoding gates determine which functions of the input the circuit can represent~\cite{schuld2021}; we use angle embedding (\cref{sec:hybrid}).

VQCs are widely studied in the noisy intermediate-scale quantum (NISQ) era~\cite{preskill2018}, i.e.\ today's small, error-prone quantum processors. Our circuits are run on classical simulators; we do not use quantum hardware. A statevector simulator represents the full quantum state as a list of $2^n$ complex numbers (amplitudes) and, in the worst case, updates the whole list for each gate. While adding a layer increases the gate count without changing the state size, adding a qubit doubles that size, creating an exponential state-size dependence on qubit count. Still, actual runtime also depends on gate count, batching and implementation overhead~\cite{xu2023}.

\subsection{How circuit gradients are obtained}\label{sec:shift}

On a classical simulator, as in our implementation, gradients are obtained by backpropagation: each backward pass retraces the operations that simulate the circuit, so the cost of every stage grows with the simulated circuit. On quantum hardware, however, this approach does not carry over directly, as it is not clear how intermediate quantum states or derivatives could be stored and reused inside a quantum computation~\cite{schuld2019}. Gradients of circuit outputs can instead be obtained using methods such as the parameter-shift rule~\cite{mitarai2018,schuld2019}: for a gate generated by a Pauli operator, the derivative with respect to its angle is computed from two runs of the circuit with that angle shifted up and down. Higher-order derivatives follow by repeating the rule, with up to $2^d$ shifted runs for each order-$d$ derivative with respect to a given set of gate angles (four for one second-order entry)~\cite{mari2021}. The same rule applies to the encoding gates, which gives derivatives of a quantum model with respect to its input~\cite{kyriienko2021}. In our hybrid network the circuit input is produced by classical layers, so input derivatives of the full network combine the chain rule through those layers with derivatives of the circuit; a second-order input derivative of the full network therefore needs many such circuit entries, not four circuit runs. We measure the simulator case only (hardware is discussed in \cref{sec:discussion}).

\subsection{Notation}

\begin{table}[H]
\centering
\caption{Notation and terminology.}
\label{tab:notation}
\begin{tabular}{>{\raggedright\arraybackslash}p{0.36\linewidth}>{\raggedright\arraybackslash}p{0.56\linewidth}}
\toprule
Symbol / term & Meaning \\
\midrule
$x = (z,t)$ & network input: depth and time \\
$\That$ & network output (temperature), with trainable parameters $\theta$ \\
$k_z$ & thermal diffusivity \\
$w_{\mathrm{PDE}}$ & weight of the PDE term in the loss (1 for Dirichlet and Neumann, 10 for Robin) \\
$\nabla_x\That$ & first-order input gradient (\firstgrad{} stage) \\
$\partial^2\That/\partial z^2$ & second-order input gradient (\secondgrad{} stage) \\
$\partial\Lpde/\partial\theta$ & parameter gradient (\paramgrad{} stage; parameter gradient of the PDE loss) \\
$n$, $L$ & number of qubits (width), number of circuit layers (depth) \\
\fwd, \firstgrad,\newline \secondgrad, \paramgrad & the four cumulative timed stages (\cref{sec:stages}) \\
\bottomrule
\end{tabular}
\end{table}

\section{Model architectures}\label{sec:models}

\subsection{Classical baseline}

The classical PINN is a fully connected network with five hidden layers of width 32 and tanh activations, mapping $(z,t)\mapsto\That$; it has 4353 trainable parameters. The classical network takes $(z,t)$ directly; the hybrid and its controls first map the input to $[-1,1]$ (\cref{sec:hybrid}). A fully connected layer with $a$ inputs and $b$ outputs has $ab+b$ trainable weights and biases. Each layer takes a list of numbers, mixes them with adjustable weights and passes the result through a smooth S-shaped function (tanh) before handing it to the next layer.

\subsection{Hybrid architecture}\label{sec:hybrid}

\begin{figure}[ht]
\centering
\includegraphics[width=\linewidth]{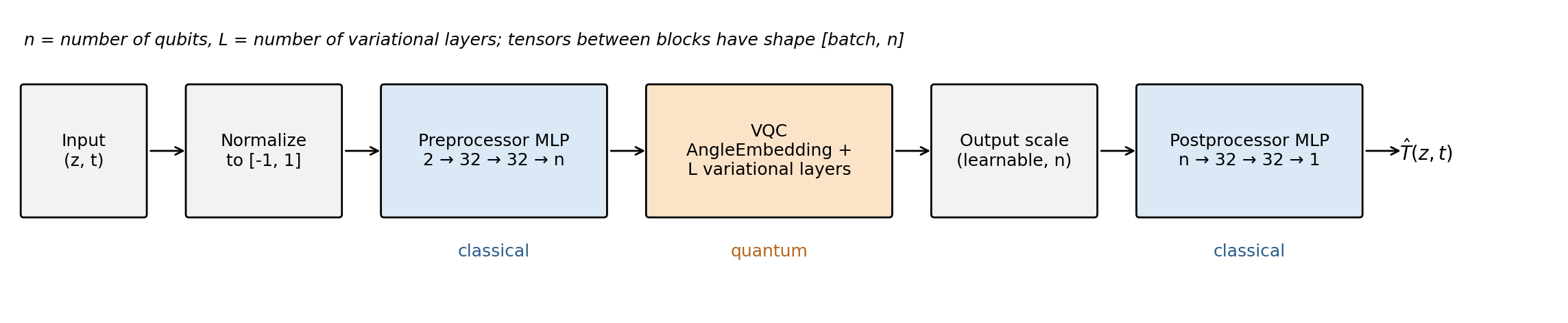}
\caption{Hybrid QCPINN: classical layers, quantum layer, classical layers.}
\label{fig:arch}
\end{figure}

The hybrid network (\cref{fig:arch}) replaces the middle of the network with a VQC. A classical pre-processor $2\to32\to32\to n$ (tanh activations) maps the normalized input to $n$ features. The VQC turns these into $n$ Pauli-$Z$ expectation values (one measured average per qubit), a learnable per-qubit scale is applied, and a classical post-processor $n\to32\to32\to1$ produces $\That$.

\paragraph{Circuit.} The features are encoded by angle embedding: each feature sets the angle of one $R_X$ rotation on its own qubit. Here $R_X$ and $R_Z$ denote rotations about the X and Z axes, and a controlled-NOT (CNOT) is a two-qubit gate used to connect qubits. Each of the $L$ layers applies $R_Z$ and $R_X$ rotations on every qubit, a ring of CNOT gates $i\to(i+1)\bmod n$ that entangles neighboring qubits, then $R_X$ and $R_Z$ on every qubit again, giving $4nL$ trainable quantum parameters. The output is $\langle Z_i\rangle$ for each qubit. The circuit is adapted from the layered ansatz of Farea et al.~\cite{farea2025}, which alternates $R_X$/$R_Z$ rotations with nearest-neighbor CNOT gates; we close the CNOT chain into a ring, so each layer has $n$ rather than $n-1$ CNOT gates. Gradients are obtained by backpropagation on PennyLane's \dqubit{} simulator~\cite{bergholm2018}.

\subsection{Configurations}

\begin{figure}[ht]
\centering
\includegraphics[width=\linewidth]{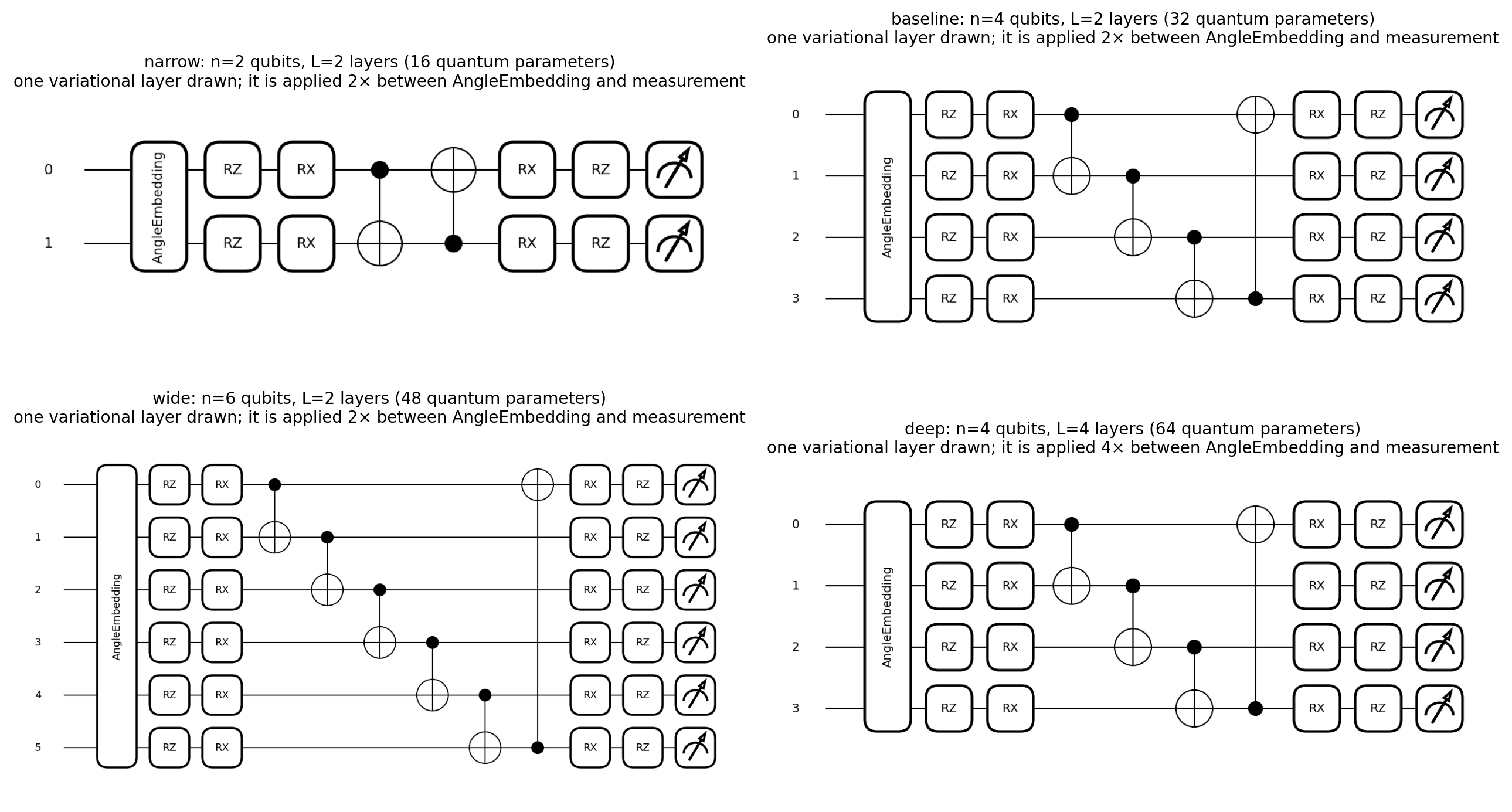}
\caption{Exemplar circuits of the four hybrid configurations (narrow, hybrid-4q, wide, deep). Each horizontal line is one qubit; boxes are rotation gates, linked dots and circles are CNOT gates, and the meter symbols on the right are measurements.}
\label{fig:circuits}
\end{figure}

\Cref{tab:configs} lists the four hybrid configurations. We call the 4-qubit, 2-layer reference hybrid \emph{hybrid-4q} (32 quantum parameters) to distinguish it from the classical baseline; the other three configurations each change one of its two circuit dimensions. \emph{Wide} increases the qubit count of the hybrid-4q from 4 to 6 at fixed depth; \emph{narrow} reduces it to 2; \emph{deep} keeps 4 qubits and doubles the depth to 4 layers. \Cref{fig:circuits} shows an exemplar circuit for each.

\begin{table}[ht]
\centering
\caption{Configurations. Quantum parameters are $4nL$; the classical share includes the $n$ learnable output scales.}
\label{tab:configs}
\begin{tabular}{lcc rrrr}
\toprule
Config & Qubits $n$ & Layers $L$ & Quantum & Classical & Total & Ratio to classical \\
\midrule
classical & --- & --- & 0 & 4353 & 4353 & 1.00 \\
narrow & 2 & 2 & 16 & 2405 & 2421 & 0.56 \\
hybrid-4q & 4 & 2 & 32 & 2537 & 2569 & 0.59 \\
wide & 6 & 2 & 48 & 2669 & 2717 & 0.62 \\
deep & 4 & 4 & 64 & 2537 & 2601 & 0.60 \\
\bottomrule
\end{tabular}
\end{table}

\subsection{Size-matched classical controls}\label{sec:controls}

For each hybrid, we time a size-matched classical control that replaces the VQC with a tanh layer of width $n$ while retaining the surrounding classical components. Its parameter count matches the hybrid's classical parameter count, not its total count. The controls are \texttt{ctrl\_n2}, \texttt{ctrl\_n4} and \texttt{ctrl\_n6}; deep and hybrid-4q share \texttt{ctrl\_n4}. This comparison measures the additional runtime associated with the simulated circuit relative to that classical replacement.

A separate \emph{classical (Adam) control} is used in \cref{sec:training}. It is the larger 4353-parameter classical PINN trained with an Adam-only schedule. Unlike the size-matched controls, it is used to interpret differences in training procedure.

\Cref{tab:comparisons} summarizes the three comparisons used in this paper and the question each one answers.

\begin{table}[ht]
\centering
\caption{The three comparisons made in this paper.}
\label{tab:comparisons}
\begin{tabular}{>{\raggedright\arraybackslash}p{0.30\linewidth}>{\raggedright\arraybackslash}p{0.48\linewidth}>{\raggedright\arraybackslash}p{0.12\linewidth}}
\toprule
Hybrid compared with & Question answered & Section \\
\midrule
Classical baseline (4353 parameters) & How do parameter count and runtime change relative to the selected classical PINN? & \ref{sec:timing}, \ref{sec:params} \\
Size-matched control (\texttt{ctrl\_n2}, \texttt{ctrl\_n4}, \texttt{ctrl\_n6}) & With the same classical layers, how much runtime does the simulated circuit add? & \ref{sec:timing} \\
Classical (Adam) control & How does the accuracy gap change when the classical PINN uses an Adam-only schedule like the hybrids? & \ref{sec:training} \\
\bottomrule
\end{tabular}
\end{table}

\section{Experimental protocol}\label{sec:protocol}

\subsection{What is timed}

We measure the four cumulative stages described in \cref{sec:stages} (\fwd, \firstgrad, \secondgrad, \paramgrad) using the same batch of collocation points. A batch is defined as the set of points processed together in a single call. By default, each batch contains 2540 collocation points sampled uniformly from $[0,2]\times[0,5]$, matching the training batch size. Batch size is also varied from 100 to 5000 in a separate experiment (\cref{sec:batch}). In addition, the circuit-only forward pass is compared across two simulator backends: \dqubit{}, which processes the entire batch in a single vectorized call, and \lqubit{} (\cref{sec:backend}).

\paragraph{Compute environment.} All timings were collected on a single Windows~11 workstation (build 26200) with a 13th Gen Intel Core i5-13400F processor (10 cores, 16 threads); no GPU was used. PyTorch used its default of 10 intra-op threads (\texttt{OMP\_NUM\_THREADS} was unset). Software versions were Python 3.14.3, PyTorch 2.11.0 (CPU build), PennyLane 0.44.1 with PennyLane-Lightning 0.44.0, and NumPy 2.4.4. Inputs, network weights, and circuit parameters were stored in single precision (float32). Each results folder records the benchmark script's version-control commit.

\subsection{Timing procedure}

Each case (configuration $\times$ seed) first makes 3 untimed warm-up calls per stage. Its number of timed repetitions is then set between 5 and 20 (at most 10 for \lqubit{}) so that its timed calls total about 20\,s; in the main section, every case reached 20. Timing proceeds in rounds, and in each round every case with repetitions left calls each of its four stages once. The first-order input gradient is always computed with \texttt{create\_graph=True}, which later stages need in order to differentiate through it; the \firstgrad{} time therefore includes building this graph. Before the first timed section, a 120\,s settle phase of untimed work brings CPU clocks and thread scheduling to a steady state. This was added after an earlier run showed the hybrid configurations running up to 9\% faster at the end than at the start, consistent with CPU warm-up rather than background load. Cases (configuration $\times$ seed) are interleaved in a shuffled order within each round, so that slow drift of the machine affects all cases alike rather than biasing whichever ran last. Network weights, batch data, and circuit parameters are redrawn for each of 3 independent seeds (0, 1, 2); a seed is the starting value of the random-number generator. We report the mean over seeds of the per-seed mean (over that seed's repetitions), and the sample standard deviation across seeds. The reported uncertainty describes variation among three seed-level averages; it is not a confidence interval.

\subsection{Drift check and run selection}\label{sec:drift}

A drift check re-times three of the main cases (classical, hybrid-4q, wide) at the end of the run. For each configuration and stage, it divides the median call time at the end by the median call time from the main measurement. The run passes if the largest deviation of these ratios from 1 is at most 10\%, counting only cells whose median in the main measurement is at least 5\,ms. The excluded cells are the four classical stages (0.5--4.1\,ms), where a few scheduling-jitter outliers shift the timings by 10--30\% while the minimum barely changes; they are still reported. The size-matched controls were not included in the drift check, so the denominators of the slowdowns in \cref{sec:timing} are checked only by their spread across seeds (\cref{sec:perconfig}).

This rule replaced an earlier one that compared per-seed means across all cells and failed all of the first five runs, mainly on the sub-millisecond cells. The new rule was fixed before the sixth run was taken; re-judged under it, runs 2, 4 and 5 pass (largest deviations 6.8\%, 6.3\% and 8.8\%) and runs 1 and 3 do not. The sixth run passes with a largest deviation of 4.2\% (wide \secondgrad), the smallest of the six, and its numbers are the ones reported in \cref{sec:timing,sec:scaling,sec:params}. It was taken on a high-performance power plan, on AC power, with background applications closed and after the settle phase above. The growth-rate fits of \cref{sec:qubits,sec:depth,sec:batch} agree within a few percent across all six runs. Within the selected run, configurations that appear in more than one section were timed separately in each. These repeated measurements differ by 1--5\% in most cases and by up to 15\% in one (narrow \firstgrad{}: $11.25$\,ms in \cref{tab:timing}, $12.89$\,ms at $n=2$ in \cref{tab:qubits}). This is larger than the spread across seeds, so the $\pm$ values understate the full measurement uncertainty, but none of the conclusions depends on differences of this size.

\section{Results I: stage-wise wall-clock cost}\label{sec:timing}

Across every stage, each hybrid is at least 10$\times$ slower than the classical PINN, and up to 97$\times$ slower for the parameter gradient used in training. All numbers are from the run selected in \cref{sec:drift}; times are in milliseconds (ms), mean over 3 seeds $\pm$ sample standard deviation across seeds, on the default batch of 2540 collocation points unless stated otherwise.

\subsection{Per-configuration timing}\label{sec:perconfig}

\Cref{tab:timing} gives the four timed stages for the classical baseline, the four hybrid configurations and their size-matched controls (\cref{sec:controls}).

\begin{table}[ht]
\centering
\caption{Stage timing at the default batch size (ms, mean $\pm$ std across 3 seeds).}
\label{tab:timing}
\begin{tabular}{lrrrr}
\toprule
Config & \fwd & \firstgrad & \secondgrad & \paramgrad \\
\midrule
classical & $0.53 \pm 0.01$ & $0.95 \pm 0.09$ & $1.61 \pm 0.01$ & $4.06 \pm 0.07$ \\
narrow & $7.60 \pm 0.13$ & $11.25 \pm 0.16$ & $21.67 \pm 0.45$ & $42.15 \pm 1.16$ \\
hybrid-4q & $16.08 \pm 0.60$ & $30.44 \pm 0.20$ & $62.65 \pm 0.84$ & $126.4 \pm 0.2$ \\
wide & $29.88 \pm 0.13$ & $80.93 \pm 1.62$ & $153.7 \pm 0.7$ & $394.3 \pm 6.2$ \\
deep & $26.34 \pm 0.47$ & $52.54 \pm 1.09$ & $97.30 \pm 1.06$ & $203.7 \pm 1.3$ \\
\midrule
\texttt{ctrl\_n2} & $0.70 \pm 0.00$ & $1.09 \pm 0.03$ & $1.90 \pm 0.01$ & $4.54 \pm 0.04$ \\
\texttt{ctrl\_n4} & $0.70 \pm 0.01$ & $1.11 \pm 0.04$ & $1.95 \pm 0.02$ & $4.58 \pm 0.04$ \\
\texttt{ctrl\_n6} & $0.71 \pm 0.01$ & $1.14 \pm 0.01$ & $2.01 \pm 0.13$ & $4.65 \pm 0.07$ \\
\bottomrule
\end{tabular}
\end{table}

\Cref{tab:slowdown} converts these into slowdowns: how many times longer a hybrid takes than the classical baseline, and than its own size-matched control. Although the classical baseline's cells, all below 5\,ms, are excluded from the drift check (\cref{sec:drift}), their spread across seeds is 1--10\% of the mean (largest for \firstgrad{}), so the slowdowns against it carry an uncertainty of similar size.

\begin{table}[ht]
\centering
\caption{Slowdown relative to the classical baseline and to each hybrid's size-matched control (narrow $\to$ \texttt{ctrl\_n2}, hybrid-4q and deep $\to$ \texttt{ctrl\_n4}, wide $\to$ \texttt{ctrl\_n6}). fwd, 1st, 2nd and param abbreviate the \fwd, \firstgrad, \secondgrad{} and \paramgrad{} stages.}
\label{tab:slowdown}
\begin{tabular}{l rrrr rrrr}
\toprule
 & \multicolumn{4}{c}{vs.\ classical} & \multicolumn{4}{c}{vs.\ matched control} \\
\cmidrule(lr){2-5}\cmidrule(lr){6-9}
Config & fwd & 1st & 2nd & param & fwd & 1st & 2nd & param \\
\midrule
narrow & 14.3$\times$ & 11.9$\times$ & 13.5$\times$ & 10.4$\times$ & 10.9$\times$ & 10.3$\times$ & 11.4$\times$ & 9.3$\times$ \\
hybrid-4q & 30.3$\times$ & 32.1$\times$ & 38.9$\times$ & 31.1$\times$ & 23.0$\times$ & 27.4$\times$ & 32.1$\times$ & 27.6$\times$ \\
wide & 56.3$\times$ & 85.4$\times$ & 95.5$\times$ & 97.1$\times$ & 42.2$\times$ & 70.7$\times$ & 76.4$\times$ & 84.9$\times$ \\
deep & 49.7$\times$ & 55.5$\times$ & 60.5$\times$ & 50.2$\times$ & 37.7$\times$ & 47.3$\times$ & 49.8$\times$ & 44.4$\times$ \\
\bottomrule
\end{tabular}
\end{table}

\paragraph{The slowdown remains large against size-matched controls.}
The hybrids remain 9.3--84.9$\times$ slower than their size-matched classical
controls, which retain the surrounding classical components but replace
the circuit with $\tanh$. One might expect this gap to disappear once the different classical architecture is accounted for; on the contrary, the additional cost persists. It is important to note, however, that these ratios characterize PennyLane's \dqubit{} implementation with backpropagation under the tested conditions.

\paragraph{The extra derivatives cost more in larger circuits.} The ratio of a later stage to \fwd{} is cumulative (\cref{sec:stages}): a ratio of 3.90 means 2.90 forward passes' worth of extra work. For \secondgrad{}, this ratio is 3.03 for the classical baseline, 2.85 for narrow, 3.90 for hybrid-4q, 3.69 for deep, and 5.14 for wide. For \paramgrad{}, it is 7.65 for classical, 5.54 for narrow, 7.86 for hybrid-4q, 7.73 for deep, and 13.20 for wide. Across the three configurations with two circuit layers, these ratios
increase with qubit count. Doubling depth at four qubits produces similar ratios, although it increases the absolute time of every stage.

\subsection{Excess time relative to size-matched controls}\label{sec:share}

To quantify the additional wall-clock cost associated with the circuit-based
implementation, we compare each hybrid with its size-matched
classical control (\cref{sec:controls}). We define the excess-time fraction as
\[
  f_{\mathrm{excess}}
  = \frac{t_{\mathrm{hybrid}}-t_{\mathrm{control}}}
         {t_{\mathrm{hybrid}}}.
\]
This is a comparison between two implementations, not a direct partition
of the hybrid's execution time into quantum and classical operations.
It includes differences in circuit execution, differentiation, and
framework overhead.

\begin{table}[ht]
\centering
\caption{Excess time relative to the size-matched classical control, expressed
as a percentage of hybrid stage time,
$100(1-t_{\mathrm{control}}/t_{\mathrm{hybrid}})$.
Values are calculated from the mean times in \cref{tab:timing}
at a batch size of 2540.}
\label{tab:qshare}
\begin{tabular}{lrrrr}
\toprule
Config & \fwd & \firstgrad & \secondgrad & \paramgrad \\
\midrule
narrow    & 91\% & 90\% & 91\% & 89\% \\
hybrid-4q & 96\% & 96\% & 97\% & 96\% \\
wide      & 98\% & 99\% & 99\% & 99\% \\
deep      & 97\% & 98\% & 98\% & 98\% \\
\bottomrule
\end{tabular}
\end{table}

With an excess-time fraction of 89--99\% across the tested configurations and stages, replacing the circuit with the classical control removes most of the measured runtime, including at the derivative stages. Timing the circuit alone gives a similar picture: for hybrid-4q at the default batch, the circuit takes $14.16$\,ms of the $16.08$\,ms forward pass (\cref{tab:timing,tab:backend}), about 88\%.

\section{Results II: scaling of wall-clock cost}\label{sec:scaling}

This section varies qubit count, circuit depth, and batch size one at a time around the hybrid-4q configuration, and compares two simulator backends.

\subsection{Qubit sweep}\label{sec:qubits}

\Cref{tab:qubits} reports timings as the qubit count $n$ increases
from 2 to 12, with circuit depth fixed at two layers and batch size
fixed at 2540. The classical bottleneck dimensions change with $n$. Gradient stages are omitted above 8 qubits, where a single forward pass already exceeds half a second and repeated backward passes become impractical to collect within the benchmark's time budget.

\begin{table}[ht]
\centering
\caption{Qubit sweep, 2 layers (ms).}
\label{tab:qubits}
\begin{tabular}{rrrrr}
\toprule
$n$ qubits & \fwd & \firstgrad & \secondgrad & \paramgrad \\
\midrule
2 & $7.44 \pm 0.09$ & $12.89 \pm 0.50$ & $21.05 \pm 0.49$ & $41.09 \pm 0.24$ \\
4 & $16.20 \pm 0.26$ & $32.35 \pm 0.34$ & $62.77 \pm 1.20$ & $127.0 \pm 0.6$ \\
6 & $31.58 \pm 0.28$ & $81.43 \pm 1.49$ & $147.6 \pm 0.5$ & $386.1 \pm 2.3$ \\
8 & $121.5 \pm 0.3$ & $332.3 \pm 3.0$ & $778.8 \pm 0.5$ & $2229.5 \pm 3.1$ \\
10 & $681.6 \pm 6.0$ & --- & --- & --- \\
12 & $3176.4 \pm 5.2$ & --- & --- & --- \\
\bottomrule
\end{tabular}
\end{table}

As the qubit count increases, time grows approximately exponentially, with the forward pass rising from 7.44\,ms at 2 qubits to 3176.4\,ms at 12 qubits, more than 400-fold. A $\log_2$-linear fit (a straight line through the logarithm of time) gives a per-qubit growth factor of 1.84$\times$ for \fwd, 1.71$\times$ for \firstgrad, 1.79$\times$ for \secondgrad{} and 1.92$\times$ for \paramgrad. The growth factors below two are consistent with a contribution from
fixed execution overhead, particularly at small $n$.

Two features of the table refine these averages. First, the \fwd{} fit spans 2--12 qubits while the gradient fits span only 2--8. Over the common range of 2 to 8 qubits, \fwd{} grows 16-fold, \firstgrad{} 26-fold, \secondgrad{} 37-fold and \paramgrad{} 54-fold, so the derivative stages grow markedly faster than the plain prediction. Second, growth is not constant. Each two added qubits multiply \fwd{} time by about 2 up to 6 qubits, where fixed overheads may contribute substantially, but by 3.8--5.6 from 6 to 12 qubits, about the doubling per qubit expected from the state size. At 8 qubits, one \paramgrad{} already takes 2229.5\,ms, about 550 times the classical baseline's 4.06\,ms.

\subsection{Circuit depth sweep}\label{sec:depth}

\Cref{tab:depth} times a 4-qubit circuit as the depth $L$ increases from 1 to 8 layers.

\begin{table}[ht]
\centering
\caption{Depth sweep, 4 qubits (ms).}
\label{tab:depth}
\begin{tabular}{rrrrr}
\toprule
$L$ layers & \fwd & \firstgrad & \secondgrad & \paramgrad \\
\midrule
1 & $10.83 \pm 0.11$ & $19.27 \pm 0.13$ & $40.18 \pm 0.30$ & $86.73 \pm 0.15$ \\
2 & $15.91 \pm 0.27$ & $30.16 \pm 0.88$ & $61.32 \pm 0.08$ & $125.4 \pm 1.2$ \\
4 & $25.86 \pm 0.52$ & $49.04 \pm 0.79$ & $94.75 \pm 0.55$ & $201.8 \pm 3.3$ \\
8 & $45.42 \pm 0.48$ & $91.28 \pm 1.05$ & $161.7 \pm 0.8$ & $354.9 \pm 2.1$ \\
\bottomrule
\end{tabular}
\end{table}

Unlike qubits, which multiply the cost, depth adds to it roughly linearly. A linear fit gives 4.94\,ms per added layer for \fwd, 10.24\,ms for \firstgrad, 17.15\,ms for \secondgrad{} and 38.30\,ms for \paramgrad. Going from 1 to 8 layers, an eightfold increase in quantum parameters, multiplies \fwd{} time by about 4.2 and \paramgrad{} time by about 4.1. Each layer adds an equal-sized block of gates to simulate and differentiate through, rather than doubling the size of the simulated state the way an added qubit does.

\subsection{Batch-size sweep}\label{sec:batch}

\Cref{tab:batch} times the hybrid-4q configuration (4 qubits, 2 layers) as the batch size increases from 100 to 5000, holding the network fixed; 2540 is the training batch size used elsewhere in this paper.

\begin{table}[ht]
\centering
\caption{Batch-size sweep, hybrid-4q configuration (ms).}
\label{tab:batch}
\begin{tabular}{rrrrr}
\toprule
Batch & \fwd & \firstgrad & \secondgrad & \paramgrad \\
\midrule
100 & $11.63 \pm 0.25$ & $16.40 \pm 0.64$ & $25.04 \pm 0.47$ & $49.74 \pm 0.85$ \\
500 & $13.56 \pm 0.18$ & $20.52 \pm 0.77$ & $34.89 \pm 0.41$ & $67.89 \pm 0.43$ \\
1000 & $14.22 \pm 0.03$ & $24.47 \pm 0.50$ & $43.38 \pm 0.44$ & $89.93 \pm 1.18$ \\
2540 & $16.03 \pm 0.08$ & $31.41 \pm 0.13$ & $63.14 \pm 1.13$ & $125.7 \pm 0.6$ \\
5000 & $18.91 \pm 0.10$ & $42.01 \pm 0.78$ & $81.60 \pm 0.51$ & $183.8 \pm 0.6$ \\
\bottomrule
\end{tabular}
\end{table}

Time grows much more slowly than batch size: a 50-fold increase in batch size, from 100 to 5000 points, raises \fwd{} time only 1.6-fold and \paramgrad{} time 3.7-fold. A log-log fit of time against batch size gives an exponent of 0.12 for \fwd{} rising to 0.33 for \paramgrad{} (an exponent of 1 would mean time proportional to batch size). This sub-linear growth is consistent with vectorized batch execution
and the amortization of execution overhead across more samples
(\cref{sec:backend}). The fitted exponent is larger for the later stages. This sweep measures hybrid runtime; the classical baseline was not
evaluated across batch sizes.

\subsection{Backend comparison}\label{sec:backend}

\Cref{tab:backend} compares the circuit-only forward pass on \dqubit{} against \lqubit{}, PennyLane's C++ statevector backend, across the four configurations and the batch sweep.

\begin{table}[ht]
\centering
\caption{Circuit-only forward pass: \dqubit{} vs.\ \lqubit{} (ms).}
\label{tab:backend}
\begin{tabular}{lrrr}
\toprule
Case & \dqubit & \lqubit & lightning / default \\
\midrule
narrow & $6.41 \pm 0.25$ & $719.7 \pm 1.7$ & 112.3$\times$ \\
hybrid-4q & $14.16 \pm 0.27$ & $1417.5 \pm 6.1$ & 100.1$\times$ \\
wide & $27.46 \pm 0.53$ & $2073.3 \pm 0.6$ & 75.5$\times$ \\
deep & $22.73 \pm 0.38$ & $2502.7 \pm 6.3$ & 110.1$\times$ \\
\midrule
batch 100 & $8.14 \pm 0.21$ & $84.23 \pm 1.66$ & 10.3$\times$ \\
batch 500 & $9.97 \pm 0.36$ & $293.1 \pm 1.8$ & 29.4$\times$ \\
batch 1000 & $11.80 \pm 0.18$ & $551.3 \pm 2.7$ & 46.7$\times$ \\
batch 2540 & $14.37 \pm 0.16$ & $1416.5 \pm 2.7$ & 98.6$\times$ \\
batch 5000 & $16.53 \pm 0.29$ & $2733.3 \pm 8.2$ & 165.3$\times$ \\
\bottomrule
\end{tabular}
\end{table}

Although \lqubit{} is a C++ backend, in this setting it is slower than \dqubit{}: 75.5--112.3$\times$ slower across the four configurations at the default batch, and from 10.3$\times$ slower at a batch of 100 to 165.3$\times$ at a batch of 5000. A log-log fit gives a batch exponent of 0.18 on \dqubit{} but 0.89, essentially linear, on \lqubit{}.

Both backends are called the same way, with the batch passed as one array to a single circuit evaluation. \dqubit{} vectorizes this call across the batch. In this execution path, \lqubit{} appears to evaluate the circuit once per sample instead: its time grows almost in proportion to batch size (exponent 0.89, against 0.18 for \dqubit{}). At the batch sizes used to train this PINN (thousands of collocation points), this per-sample loop is the main reason \lqubit{} is slower. Gate count still matters, though: on \lqubit{}, deep (4 qubits, 4 layers) is slower than wide (6 qubits, 2 layers), the opposite of their order on \dqubit{}.

\section{Results III: parameter count versus wall-clock cost}\label{sec:params}

Having compared the models by wall-clock time in \cref{sec:timing,sec:scaling}, we now ask: would their parameter counts have predicted those differences? The hybrids have 37.6--44.4\% fewer parameters than the classical baseline, but the quantum circuit contributes at most 2.46\% of their total. \Cref{tab:configs} (\cref{sec:models}) lists each configuration's classical and quantum parameter count; \cref{tab:share} restates them as the two quantities relevant to the parameter-count claim.

\begin{table}[ht]
\centering
\caption{Quantum share of the total parameter count, and the reduction relative to the classical baseline (4353 parameters). Quantum share is $4nL$ divided by total parameters.}
\label{tab:share}
\begin{tabular}{lrr}
\toprule
Config & Quantum share & Reduction vs.\ classical \\
\midrule
narrow & 0.66\% & 44.4\% \\
hybrid-4q & 1.25\% & 41.0\% \\
wide & 1.77\% & 37.6\% \\
deep & 2.46\% & 40.2\% \\
\bottomrule
\end{tabular}
\end{table}

Across the four hybrid configurations, the VQC itself contributes only 16--64 trainable parameters. Total parameter count nonetheless falls by 37.6--44.4\%, because the classical pre- and post-processor around the VQC (2405--2669 parameters) is already much smaller than the classical baseline's five hidden layers of width 32 (4353 parameters).

\begin{figure}[ht]
\centering
\includegraphics[width=0.8\linewidth]{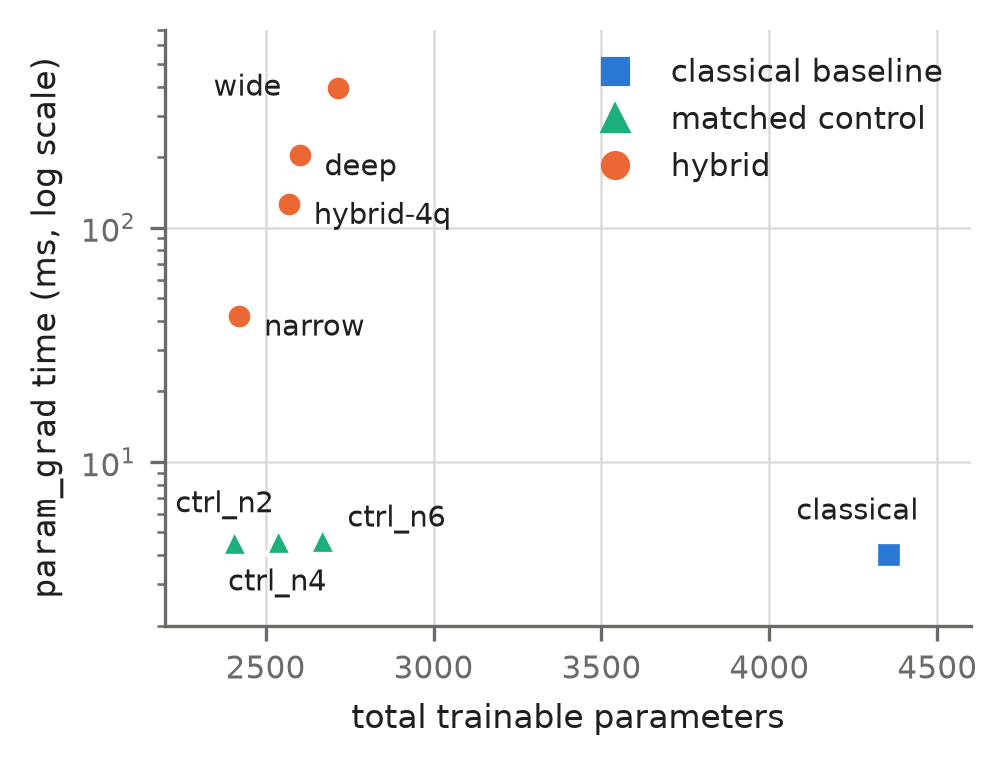}
\caption{Total trainable parameter count versus cumulative PDE-loss
parameter-gradient evaluation time at a batch size of 2540.
Timings are means across three seeds.
Parameter counts and timings are from
\cref{tab:configs,tab:timing}.}
\label{fig:params-runtime}
\end{figure}

\Cref{fig:params-runtime} shows that models with fewer parameters do not
necessarily execute faster. The size-matched classical controls and the
hybrid configurations occupy similar parameter-count ranges but have
markedly different PDE-loss parameter-gradient times.

While the reduction in total parameter count reflects the smaller classical architecture surrounding the circuit, the hybrids remain 9.3--84.9$\times$ slower than the size-matched classical controls (\cref{sec:perconfig}), so the smaller parameter count does not translate into lower wall-clock cost.

\paragraph{Fewer parameters did not make even the classical networks faster.} The size-matched controls have 2405--2669 parameters against 4353 for the classical baseline, yet they are slower at every stage: 0.70--0.71\,ms against 0.53\,ms for \fwd, and 4.54--4.65\,ms against 4.06\,ms for \paramgrad. This is why the slowdowns against the controls in \cref{tab:slowdown} are smaller than those against the classical baseline. For wide, the ratio of 97.1$\times$ to 84.9$\times$ (1.14) matches \texttt{ctrl\_n6} taking 1.15 times as long as the classical baseline (4.65\,ms versus 4.06\,ms). Both networks have six linear layers; the controls additionally normalize the input and apply the per-qubit output scale, which the classical baseline does not. A likely reason is that at 0.5--5\,ms, runtime depends more on the number of operations than on their size, so these extra elementwise operations cost more than the parameters saved; we did not time a control without them to check this. At this scale, parameter count is a poor guide to runtime even for purely classical networks.

\paragraph{More quantum parameters do not mean more time.} Although deep has the most quantum parameters (64), it is faster than wide (48) at every stage, taking 203.7\,ms against 394.3\,ms for \paramgrad. This is because circuit width sets the cost, with added qubits costing far more than added layers (\cref{sec:qubits,sec:depth}).

\section{Training under the selected optimization schedules}\label{sec:training}

As a supplementary check, this section tests whether the hybrids train to useful accuracy. Each of the five models of \cref{tab:configs}, together with the classical (Adam) control introduced below, is trained once (seed 0) on six scenarios of one PDE family: three boundary conditions, each as a forward and an inverse problem.

\paragraph{Setup.} Each scenario has a budget of $N_A$ Adam steps and $N_L$ L-BFGS iterations: $N_A = 5000$, $N_L = 10000$ for Dirichlet and Neumann, and $N_A = N_L = 20000$ with the PDE residual weighted by 10 for Robin (\cref{sec:problem}).

Adam~\cite{kingma2015} takes many small steps, each guided by the current gradient and a running average of past gradients. L-BFGS~\cite{liu1989} is a quasi-Newton method: it also estimates the curvature of the loss, and at each iteration tries several step lengths along its chosen direction (a line search) to find a good one. L-BFGS often converges much further in fewer iterations, but every trial step costs an extra loss evaluation.

The classical PINN runs $N_A$ Adam steps at learning rate $10^{-3}$ (the learning rate sets the step size) and then L-BFGS with a strong-Wolfe line search for at most $N_L$ iterations. The hybrids skip L-BFGS, whose line search needs many extra circuit evaluations per iteration, and instead run Adam for $N_A + N_L$ steps (learning rate $5\times10^{-3}$ for the circuit parameters and $10^{-3}$ for the rest) with cosine decay~\cite{loshchilov2017} to $10^{-5}$, a schedule that gradually shrinks the learning rate. The optimizer therefore differs between the two families, so a gap between a hybrid and the classical PINN reflects both the model and its optimizer.

To separate the two, we also train the classical PINN with the hybrids' schedule: Adam for $N_A + N_L$ steps at learning rate $10^{-3}$ with the same cosine decay, and no L-BFGS. We call this the classical (Adam) control. It shares the hybrids' optimizer family, number of Adam steps, and cosine decay; the only difference in the schedule is that the hybrids' circuit parameters use their own learning rate of $5\times10^{-3}$. This narrows, but does not remove, the training-procedure gap. All hybrids share one optimization scheme, and within a scenario every model uses the same seed, so all of them see the same collocation points, initial data, and observations. Accuracy is measured against the reference solution as mean absolute error (MAE), the average size of the difference between predicted and true temperature.

\begin{figure}[ht]
\centering
\begin{subfigure}{0.32\linewidth}\includegraphics[width=\linewidth]{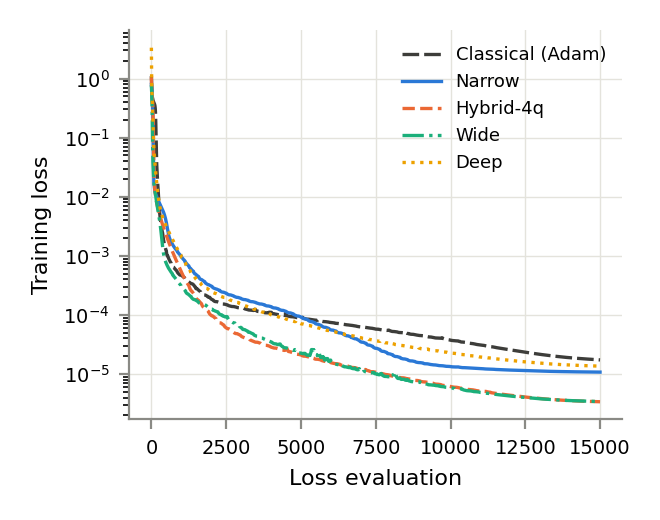}\caption{Dirichlet, forward}\end{subfigure}\hfill
\begin{subfigure}{0.32\linewidth}\includegraphics[width=\linewidth]{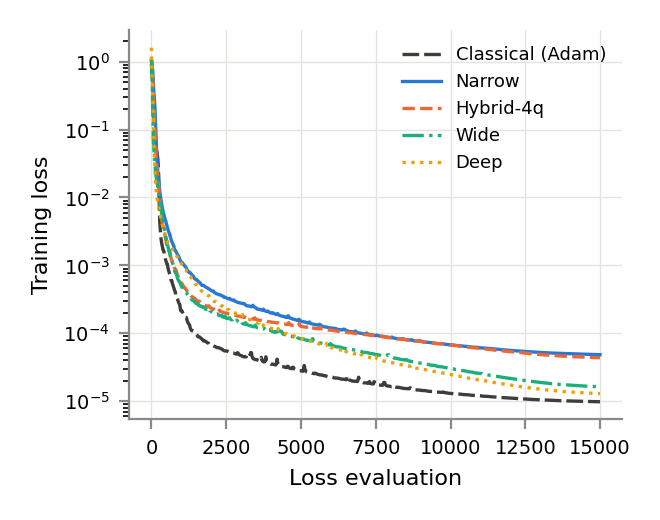}\caption{Neumann, forward}\end{subfigure}\hfill
\begin{subfigure}{0.32\linewidth}\includegraphics[width=\linewidth]{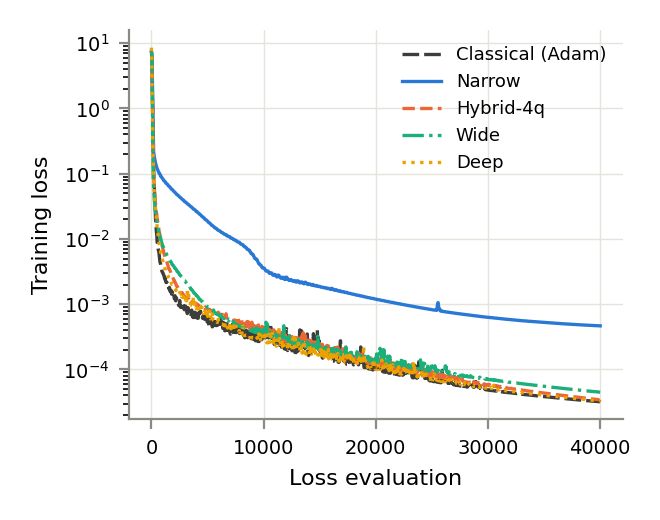}\caption{Robin, forward}\end{subfigure}\\[4pt]
\begin{subfigure}{0.32\linewidth}\includegraphics[width=\linewidth]{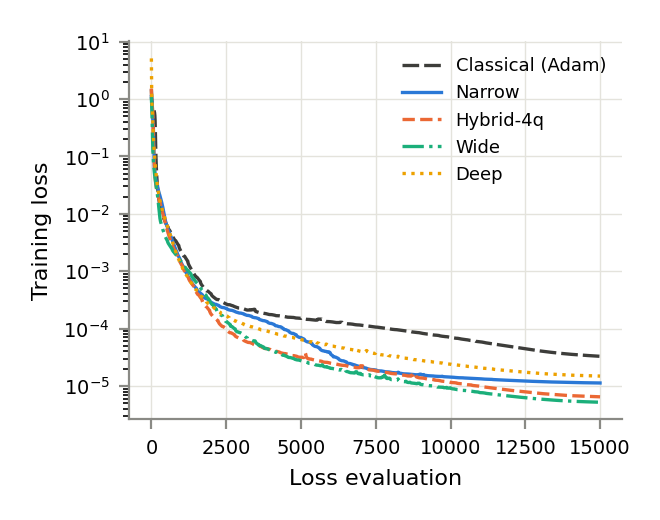}\caption{Dirichlet, inverse}\end{subfigure}\hfill
\begin{subfigure}{0.32\linewidth}\includegraphics[width=\linewidth]{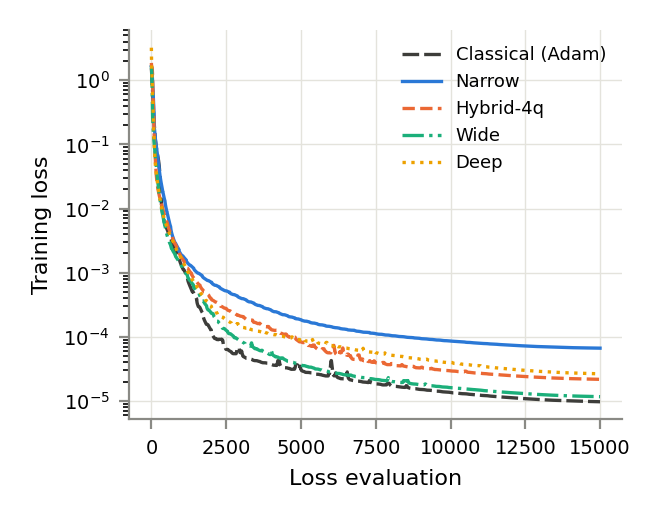}\caption{Neumann, inverse}\end{subfigure}\hfill
\begin{subfigure}{0.32\linewidth}\includegraphics[width=\linewidth]{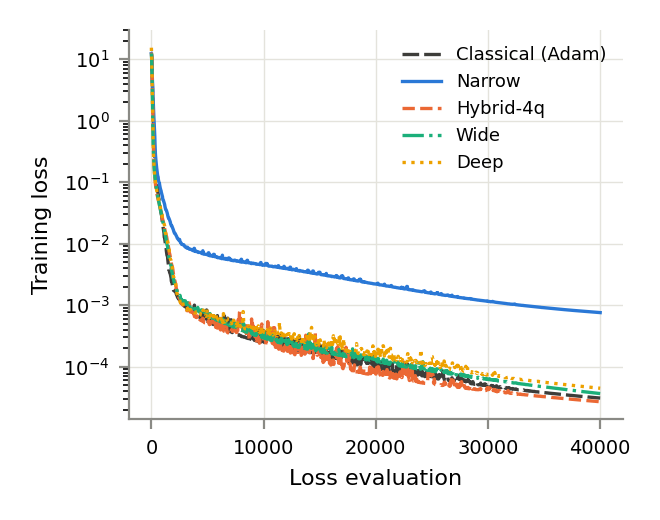}\caption{Robin, inverse}\end{subfigure}
\caption{Training loss against loss evaluations for the six scenarios (seed 0), for the classical (Adam) control and the four hybrids, all trained with Adam and cosine decay only. The classical PINN trained with Adam then L-BFGS is not shown; \cref{tab:mae} gives its test MAE. Curves are centered rolling medians over 201 loss evaluations.}
\label{fig:loss}
\end{figure}

\paragraph{Loss curves.} \Cref{fig:loss} shows the training loss of the classical (Adam) control and the four hybrids against the number of loss evaluations; all five run a fixed $N_A + N_L$ Adam steps. The classical PINN trained with L-BFGS is not plotted, because its curve would mix model and optimizer; its final accuracy is reported in \cref{tab:mae}. Curves are centered rolling medians over 201 loss evaluations, because Adam's raw loss fluctuates by orders of magnitude from step to step.

\paragraph{Accuracy.} As reported in \cref{tab:mae}, no hybrid achieves the error of the classical PINN trained with Adam followed by L-BFGS in any scenario. Against the classical (Adam) control, however, hybrid-4q's MAE is more than 10\% lower in three of six scenarios and wide's in two; in the other scenarios the difference is under 10\% or the control is more accurate. With a single seed, differences of a few percent should not be read as a ranking. Also, the hybrids' classical layers differ from those of the classical PINN, so these comparisons do not isolate what the circuit contributes to accuracy. The size-matched controls would, but they were only timed, not trained. For context, hybrid-4q's PDE-loss parameter-gradient evaluation costs
31.1$\times$ as much as that of the classical baseline
(\cref{tab:slowdown}).

\begin{table}[ht]
\centering
\small
\setlength{\tabcolsep}{3pt}
\caption{Test MAE after training (seed 0). Lower is better. Classical uses Adam then L-BFGS; the classical (Adam) control and all hybrids use Adam with cosine decay only. Bold marks the most accurate hybrid in each row; an underline marks hybrids whose MAE is more than 10\% below that of the classical (Adam) control.}
\label{tab:mae}
\begin{tabular}{l rr rrrr}
\toprule
 & \multicolumn{2}{c}{classical} & \multicolumn{4}{c}{hybrid} \\
\cmidrule(lr){2-3}\cmidrule(lr){4-7}
Scenario & Adam+L-BFGS & Adam & narrow & hybrid-4q & wide & deep \\
\midrule
Dirichlet, forward & \sci{2.31}{-4} & \sci{5.43}{-4} & \sci{8.06}{-4} & \underline{\textbf{\sci{3.40}{-4}}} & \underline{\sci{4.26}{-4}} & \sci{9.86}{-4} \\
Neumann, forward & \sci{1.66}{-4} & \sci{6.93}{-4} & \sci{1.85}{-3} & \sci{2.39}{-3} & \sci{1.31}{-3} & \textbf{\sci{1.09}{-3}} \\
Robin, forward & \sci{8.96}{-5} & \sci{4.06}{-4} & \sci{2.03}{-3} & \textbf{\sci{3.84}{-4}} & \sci{4.88}{-4} & \sci{4.23}{-4} \\
Dirichlet, inverse & \sci{2.87}{-4} & \sci{6.95}{-4} & \sci{7.13}{-4} & \underline{\textbf{\sci{5.00}{-4}}} & \sci{6.76}{-4} & \sci{1.16}{-3} \\
Neumann, inverse & \sci{2.49}{-4} & \sci{6.66}{-4} & \sci{2.80}{-3} & \sci{1.18}{-3} & \textbf{\sci{1.05}{-3}} & \sci{1.70}{-3} \\
Robin, inverse & \sci{8.66}{-5} & \sci{3.42}{-4} & \sci{2.34}{-3} & \underline{\sci{2.90}{-4}} & \underline{\textbf{\sci{2.72}{-4}}} & \sci{6.47}{-4} \\
\bottomrule
\end{tabular}
\end{table}

\begin{table}[ht]
\centering
\small
\setlength{\tabcolsep}{3pt}
\caption{Absolute error of the inferred diffusivity, $|\hat k_z - k_z|$ with $k_z = 0.05$, in the inverse problems (seed 0).}
\label{tab:kz}
\begin{tabular}{l rr rrrr}
\toprule
 & \multicolumn{2}{c}{classical} & \multicolumn{4}{c}{hybrid} \\
\cmidrule(lr){2-3}\cmidrule(lr){4-7}
Scenario & Adam+L-BFGS & Adam & narrow & hybrid-4q & wide & deep \\
\midrule
Dirichlet & \sci{8.16}{-6} & \sci{2.92}{-5} & \sci{6.13}{-5} & \sci{4.48}{-5} & \sci{2.37}{-5} & \sci{5.56}{-5} \\
Neumann   & \sci{2.56}{-5} & \sci{1.04}{-4} & \sci{6.03}{-5} & \sci{1.67}{-5} & \sci{8.98}{-5} & \sci{2.68}{-4} \\
Robin     & \sci{6.46}{-9} & \sci{1.94}{-5} & \sci{1.39}{-4} & \sci{3.63}{-6} & \sci{9.49}{-6} & \sci{3.21}{-5} \\
\bottomrule
\end{tabular}
\end{table}

That said, no hybrid configuration is best everywhere, although hybrid-4q or wide is the most accurate hybrid in five of the six scenarios, with deep leading in the other. Going from narrow to hybrid-4q (2 to 4 qubits) usually helps; more qubits or more depth beyond that do not reliably help further, and narrow and deep are never more accurate than the classical (Adam) control. \Cref{tab:kz} shows that all models recover the diffusivity to within 0.6\% of its true value. The ranking here differs from the MAE ranking: in the Neumann problem, for example, hybrid-4q has a larger MAE than the classical (Adam) control but a smaller diffusivity error.

\paragraph{Loss evaluations and wall-clock time.} The classical PINN's L-BFGS phase meets its stopping tolerance well inside its budget in every scenario, after 500 to 1700 loss evaluations, so it evaluates the loss 5500 to 6200 times in total (Dirichlet and Neumann) and about 21000 to 22000 times (Robin). Each hybrid evaluates it a fixed $N_A + N_L$ times, 15000 and 40000 respectively, i.e.\ 1.8 to 2.7 times as often as the classical PINN.

Under an equal budget of 15000 evaluations, the controlled
\paramgrad{} timings imply approximately 61\,s for the classical
model and 31.6\,min for hybrid-4q. These estimates assume constant per-evaluation cost
and cover only the PDE-loss parameter-gradient computation.
They exclude the other loss terms and optimizer updates.
Measured end-to-end training times are reported below.

Training a hybrid took from about 15 minutes (narrow, Dirichlet) to over 5 hours (wide, Robin forward), against 35 seconds to under 6 minutes for the classical PINN. These single runs on a shared workstation are too noisy for slowdown factors. Background load slowed down the batch of runs that included wide's inverse problems: in that batch, the classical PINN took 571\,s and 2150\,s on the Neumann and Robin inverse problems, against 36--166\,s in the other batches, and wide's inverse problems took 2.0 to 2.7 times as long as its forward problems with the same step counts (about 1.15 times for narrow). We therefore leave those runs out of the ranges above. \Cref{sec:timing} gives the controlled per-evaluation costs.

\section{Discussion}\label{sec:discussion}

\paragraph{Parameter count is not a substitute for wall-clock measurement.} The hybrids use 37.6--44.4\% fewer parameters than the classical PINN, yet their PDE-loss parameter-gradient evaluation is 10.4--97.1$\times$ slower (\cref{sec:perconfig}). More importantly, when compared with the size-matched controls, which share the hybrids' classical layers, the hybrids remain 9.3--84.9$\times$ slower. Put simply, parameter count does not predict cost: despite having 39--45\% fewer parameters than the classical baseline, the controls run slower at every stage, and although deep has more quantum parameters than wide, it runs faster because qubits, rather than parameters, set the size of the simulated state (\cref{sec:params}). Studies of hybrid quantum PINNs should therefore report wall-clock time per stage alongside parameter count, and compare against a classical control with the same classical layers.

\paragraph{Wall-clock cost depends on what is measured and how it is executed.} The slowdown changes with the stage: 14--56$\times$ for a forward pass against 10--97$\times$ for the parameter gradient, and between 2 and 8 qubits the parameter gradient grows 54-fold while the forward pass grows 16-fold (\cref{sec:qubits}). Because a PINN needs higher-order input derivatives, a study must state which stage it times, as a slowdown measured on prediction alone does not carry over to the derivative stages. Over the tested ranges, runtime showed an approximately exponential trend with qubit count, an approximately linear trend with depth, and sub-linear growth with batch size (\cref{sec:scaling}). The execution path matters as much as the circuit: with the batch passed as one array, \lqubit{} was 10.3--165.3$\times$ slower than \dqubit{} on the circuit forward pass, so one comparison under one calling pattern cannot rank simulators in general (\cref{sec:backend}). These dependencies are consistent with the spread of earlier reports. Farea et al.~\cite{farea2025} report a 15- to 40-fold increase per training iteration, and our \paramgrad{} slowdowns for narrow (10.4$\times$) and hybrid-4q (31.1$\times$) fall in or near that range; wide (97.1$\times$) reaches the two orders of magnitude stated by Leong et al.~\cite{leong2025}; Trahan et al.~\cite{trahan2024} report nearly four times the classical time, well below all our configurations. As such, a single aggregate figure cannot show which of these factors produced it.

\paragraph{Consequences for training in this study.} L-BFGS needs an extra loss evaluation for each line-search trial, and each evaluation runs the simulated circuit, so we trained the hybrids with Adam only, while the classical PINN used Adam followed by L-BFGS and was more accurate in every scenario (\cref{sec:training}). In the single-seed Adam-only runs, hybrid-4q's MAE was more than 10\% below that of the classical (Adam) control in three of six scenarios. Hybrid training with
L-BFGS and accuracy comparisons with the size-matched controls
remain untested.

\paragraph{Limitations.} Lastly, it should be acknowledged that this study uses a single one-dimensional PDE, a CPU-only machine and one circuit design, and measures simulation only. Quantum hardware avoids explicit statevector simulation, but derivative
evaluation and finite-shot measurement introduce different costs
(\cref{sec:shift}); the present timings do not quantify those costs. Training was run with a single seed on a machine that was not reserved for the benchmark, so end-to-end training times are indicative only. The classical baseline was not timed across batch sizes. \lqubit{} was compared only on the circuit forward pass, not with its own adjoint gradient method, and GPU simulators such as that of Chen et al.~\cite{chen2026} were not tested. All timings also use float32 and PyTorch's default of 10 threads, and neither was varied.

\section{Conclusion}\label{sec:conclusion}

To conclude, this study set out to ask whether the smaller parameter counts reported for hybrid quantum PINNs carry over to wall-clock time, and on the benchmark tested here, they did not. While the hybrids did use fewer parameters, these savings came almost entirely from the smaller classical layers surrounding the circuit, and it was the simulation and differentiation of the circuit itself that took most of the time; even against classical networks with the same layers, every hybrid remained about ten times slower or more. Further, this cost depended on both what was measured and how it was executed, rising much faster with qubit count than with depth, differing between the forward pass and the derivative stages a PINN needs, and changing by one to two orders of magnitude with the simulator's calling pattern. As such, studies of hybrid PINNs should time each stage separately, compare against a classical network with the same classical layers, and report the simulator setup, so that parameter savings and runtime can be judged together. Ultimately, whether these findings hold on GPU simulators and quantum hardware, with second-order optimizers, and for other PDEs remains a question for future work.

\section*{Code and data availability}
The code and data used in this study are available from the authors on request.

\end{document}